# The Heritage Digital Twin: a bicycle made for two. The integration of digital methodologies into cultural heritage research

**Franco Niccolucci*[1], Béatrice Markhoff[2], Maria Theodoridou[3], Achille Felicetti[1], Sorin Hermon[4]**

[1] PIN, Servizi Didattici e Scientifici per l'Università di Firenze, Piazza Ciardi 25, I-59100 Prato, Italy
[2] Université de Tours, 64 Avenue Jean Portalis, F-37200 Tours, France
[3] FORTH, N. Plastira 100, Vassilika Vouton, GR-70013 Heraklion, Crete, Greece
[4] STARC-The Cyprus Institute, 20 Konstantinou Kavafi Street, CY-2121 Aglantzia, Nicosia, Cyprus

**\*Corresponding author:** franco.niccolucci@gmail.com

## Abstract

The paper concerns the definition of a novel ontology for cultural heritage based on the concept of digital twin. The ontology, called Heritage Digital Twin ontology, is a compatible extension of the well-known CIDOC CRM ISO standard for cultural heritage documentation and incorporates all the different documentation systems presently in use for cultural heritage documentation. In the authors' view, it supports documentation interoperability at a higher level than the ones currently in use and enables effective cooperation among different users.

## Plain Language Summary

The paper addresses the issue of managing the digital documentation of cultural heritage in an effective way. For this purpose, it introduces the concept of heritage digital twin, which is inspired by the similar one used in the industry and in other domains. A digital twin is the digital replica of a real-world object. It includes all the necessary information and is able to simulate - in a digital environment - the characteristics and the behaviour of its real counterpart. Two examples are fully developed to demonstrate the value of this novel approach. This research was carried out in the framework of the EU-funded projects ARIADNEplus and 4CH, both concerning the use of digital technology to study and preserve cultural heritage.

## 1. Introduction

The title of this paper is inspired by the similar one of a paper by Pollard and Bray [1] concerning how the contribution of archaeological sciences should be integrated into the process of archaeological interpretation. According to the authors, such integration is like riding a bicycle made for two, also known as a tandem. This kind of vehicle requires a strong collaboration between the two riders to pedal synchronically and the one in front must be able and willing to drive the tandem towards a common destination, on which both riders agree. The structure of the bicycle should suit a diversity of users: tall and short; married couples and perfect strangers; sportspeople and lazy ones. The way it can be used must adapt to any kind of road, dirt trails and urban well-paved streets alike.

Cycling metaphors aside, the convergence and integration of two different disciplines puts requirements to the method and the attitude of both and of all participants. Each party involved must consider the other one's perspective and adapt its own accordingly. A similar situation takes place about the interdisciplinary approach known as digital cultural heritage. Adopting such a collaborative attitude by heritage professionals and institutions is the nucleus of what is usually called digital transformation of cultural heritage. The upskilling of professionals and the allocation of resources to information technology by heritage managers is of course required, but without such change of perspective they might be ineffective.

In the present paper we report the results of research triggered by the activities carried out within two European highly interdisciplinary projects, which probably extend well beyond the project objectives.

The first project is the ARIADNE group of projects comprising ARIADNE (Advanced Research Infrastructure for Archaeological Data Networking in Europe), an FP7 project operating from 2013 to the end of 2016 [2], and its continuation ARIADNEplus[1], an H2020 project operating from 2018 until the end of 2022. The two projects are collectively indicated as ARIADNE.

The original ARIADNE objective was stated as "To turn the sparse existing archaeological repositories, into a pan-European Integrated Research Infrastructure, with easily available and harmonised access, responding to the demand of the archaeological research community of the European Research Area and beyond" while the ARIADNEplus one is simply "Extending and focusing ARIADNE". The break between the two projects, determined by the schedule of EU calls, was bridged maintaining the main project result, the ARIADNE catalogue, and preparing for the continuation.

The most visible part of the project outcomes is the ARIADNE catalogue, also called the ARIADNE Knowledge Base, accessible from the ARIADNE portal[2], with various searching options including time, location and content of the catalogued datasets. ARIADNE aggregated more than 2,000,000 archaeological datasets from a large part of Europe. ARIADNEplus has expanded ARIADNE in geographic coverage, now extending to a larger part of Europe and including datasets from Japan, Israel, Africa, USA and Latin America; and in content type, now including many more datasets concerning archaeological sciences, images, 3D models, archaeological GIS, and so on. The total number of items in the ARIADNEplus version of the catalogue is more than 3,300,000 aggregated datasets, with another hundred thousand datasets still in the aggregation queue. Such results have raised the enthusiasm of the archaeological research community as witnessed in two publications, *The ARIADNE Impact* [3] concerning ARIADNE and a forthcoming one about the added value of ARIADNEplus. This has created a vibrant community that has been nurtured in parallel by training and dissemination activities.

Such achievements have been attained thanks to research results that in ARIADNE operate behind the scenes. There are two main strands: information technology, supporting all the steps of aggregation, and providing the technical instruments to make data FAIR (Findable, Accessible, Interoperable and Reusable); and advanced work in semantics, leading among others to the development of CRMarchaeo[3], an archaeological extension of the well-known CIDOC CRM standard ontology for cultural heritage[4]. As it is well known, CIDOC CRM – also mentioned in brief as the CRM – stands for Conceptual Reference Model supported by CIDOC, the International Committee for Documentation of ICOM, the International Council of Museums and since December 2006 is an official ISO standard (ISO 21127:2014).

The ARIADNE research achievements form one of the pillars supporting the work presented in this paper.

The ARIADNE Knowledge Base is based on a subset of the CRM ontology, to which the metadata schemas of the catalogued items are mapped. The project also developed application profiles for specific subdomains such as epigraphy, scientific analyses, excavations and so on. Thus, it suggested the idea of an overarching ontology as the one presented in the present paper.

The second project supporting the present paper is the H2020 project 4CH – Competence Centre for the Conservation of Cultural Heritage[5]. This ongoing project addresses the conservation of monuments and sites with digital methods, in view of the creation of a European Competence Centre on the subject, which the project is in charge of designing. Among others, the project aims at setting up a cloud-based Knowledge Base with conservation data. This aspect has a key but not exclusive role in the project work plan. With regard to the present paper, 4CH has brought the need of extending the scope of semantic research beyond archaeology to include tangible cultural heritage as well as its intangible component. Within 4CH, implementing a suitable documentation system is still ongoing work. We anticipate that the ontology presented here will be adopted in 4CH to build its Knowledge Base.

Thus, the semantic team of both projects has been induced to start conceiving an overarching paradigm that could include the digital representation of all heritage assets, tangible or intangible alike, for any purpose.

---

Illustrating this approach and its main concept, the Heritage Digital Twin, forms the body of the present paper.

In the rest of this paper, Section 2 is a summary of the cultural heritage digital documentation's history with its current state of the art. Section 3 motivates the need of a heritage digital twin ontology. Section 4 brings the basics of the heritage digital twin concept and shows how it goes further than 3D-based data models. Section 5 introduces the overall methodology and main concepts of the Heritage Digital Twin (HDT) Ontology, and more technical descriptions are provided in Section 6 and Section 7. Examples of digital documentations pertaining to the same heritage digital twin are developed in Section 8, one for scientific analyses and another one for art history analyses, performed on the same heritage entity. The section also includes an example of use for built heritage.

## 2. Background: the state of the art in the digital documentation of cultural heritage

An even concise description of how the digital revolution of the last part of the XX century affected also documenting cultural heritage would probably require an encyclopaedia. For example, Google scholar lists about 1,900,000 papers under the search key "digital documentation of cultural heritage" and it is likely that titles wrongly included in the search results are less than titles wrongly excluded. For instance, does Google AI consider "digital archaeology" as also pertaining to "digital documentation of cultural heritage"?

Archaeology started to be digitally documented in the second half of 1900 using relational databases, while in general similar IT applications to cultural heritage were less frequent at that time. We will follow the development of digital archaeological documentation as an indicator of the progress of such techniques.

At the beginning, digital methods were also called computer applications or quantitative methods, hence the name of the most important conference on the subject: Computer Applications and quantitative methods in Archaeology, CAA[6]. This annual event has been for a long time the main place to communicate such applications, together with other initiatives such as the dedicated journal "*Archeologia e Calcolatori*" (*Archaeology and Computers*, a vintage name indeed), which started publications in 1990 and is still published in Italy. This journal published in 2019 [4, 5] a historical account of the development of computing applications, their perspectives and their impact on archaeological theory and methods. Many papers published in the CAA proceedings in the last decade of the XX century describe examples of database use to digitally store archaeological data. The number of such applications to individual case studies continued to increase until such applications became standard practice and developing a database to collect the data of a specific archaeological investigation was no more a contribution to advancing research in this field, but just another example. At this point, the annual number of related papers collapsed. In parallel, the interest on GIS (Geographic Information Systems) increased, as it appeared that this technique might capture the locational characteristic and spatial relationships of archaeology. Non-canonical threads as the system proposed in 1979 by Jean-Claude Gardin denominated *Logicisme* [6] received less favour possibly also because they were out of the current mainstream in digital archaeology. The interest in GIS was also an obstacle to acceptance, as they need only very simple databases to document archaeological assets and related spatial information.

Two recent papers, one by Dallas [7] and the other one by Moscati [4] have brought Logicisme on the forefront again, and an excavation documented according to this system has recently been published [8, 9]. In the meanwhile, the concept of Linked Open Data has made its way into archaeology, starting from the first decade of the XXI century. In the same period, semantic concepts started to be applied to archaeological documentation and domain ontologies showed up.

This very summary sketch of the archaeological computing history explains why it is now necessary to propose new semantic tools for the archaeological documentation, based on accepted domain standards to support interoperability. This principle has been considered in ARIADNE, where a set of common aggregation metadata are used for all datasets.

Another important thread in archaeological documentation is represented by 3D models. Initially used for communication purposes, to facilitate visitors' understanding of complex archaeological sites – an application still alive today – it was realised that such 3D models might also serve for a more complete

---

information about archaeological assets, both objects and remains. A recent book [10] combines such a 3D approach with GIS. This thread may be considered as part of the general interest for using 3D as a documentation tool in Cultural Heritage.

On the general cultural heritage side, digital documentation was limited to inventories for a long period. The turning point coincided with the definition of the CIDOC CRM[7].

The CRM is the culmination of more than two decades of standards development work by the International Committee for Documentation (CIDOC) of the International Council of Museums (ICOM). Work on the CRM itself began in 1996 under the auspices of the ICOM-CIDOC Documentation Standards Working Group. Since 2000, development of the CRM has been officially delegated by ICOM-CIDOC to the CIDOC CRM Special Interest Group (SIG). The SIG, in turn, collaborates with the ISO working group ISO/TC46/SC4/WG9 to bring the CRM to the form and status of an International Standard. This set of collaborations has resulted in the production of ISO21127:2004 and ISO21127:2014, the ISO standard editions of the CIDOC CRM.

The CRM was initially conceived for museum applications but then it gradually evolved into a general-purpose cultural heritage documentation system. The CRM is an extendable system, enabling to develop extensions compliant with the CRM to consider the diverse aspects of cultural heritage, such as monuments, sites, movable or immovable assets, tangible assets in general and finally intangible ones. There are several ongoing developments of CRM compatible models[8] covering different specialised fields

Additionally, 3D models have gradually assumed an increasing importance in cultural heritage. 3D models are yet another digital representation for museum objects, but their importance is especially increasing for architectural heritage, mainly due to the architects' practice. Some actually consider the 3D model as the root and build around it the documentation system. This explains why BIM (Building Information Modelling) is being increasingly proposed as the cornerstone of such systems. BIM is a methodology in use to design new buildings which incorporates in the building project all the necessary information about services such as water, electricity, heating networks, and more. It develops a very simple semantic of objects and of their parts, to be able to avail of graphical libraries in the 3D model construction where the necessary information such as material, characteristics, industrial producer (in new buildings) and assembling/construction phases is attached to the building parts, as walls, pillars, floors, ceilings, and so on. BIM models are based on an industry standard, IFC[9] (Industry Foundation Classes), a standardized, digital description of the built asset industry, and an open, international standard ISO 16739-1:2018. A pillar of a BIM model is the 3D representation of the building as a wireframe/solid model, i.e. a skeletal representation of it as produced by well-known tools, among others by Autocad. It consists of various points, arcs, lines, circles, and curves to clearly denote object edges and depth. Information about the various components – structured according to the IFC standard – is attached to this model. The application of BIM to cultural heritage has led to HBIM, standing for Heritage BIM. An extensive survey of HBIM applications is provided by [11], while [12] focuses on applications to archaeology.

In parallel, there has been a number of proposals to use a point-cloud 3D model as the graphical support and add annotations to it concerning various aspects of the object. This kind of application is usefully applied to conservation and restoration. Among others, it is worth quoting the model used for the restoration of the Nettuno fountain in Bologna [13] and the one currently in use for the Notre Dame restoration [14]. A recent paper proposes to reconcile HBIM and point-cloud models via AI [15].

In conclusion, heritage documentation based on 3D models, either point-cloud ones or BIM-based ones, considers the volume and the shape of the heritage asset to be documented as the root of all the documentation systems. This approach does not allow – at least so far – to consider a large part of documentation that cannot be appended to any physical component and severely limits the interoperability of such documentation systems. Documentation based on the shape/volume of assets generates the fragmentation of the documentation system, as each item is documented separately. It does not allow searching across different assets, for example comparing materials, restoration techniques and so on. Also the intangible component of heritage is generally overlooked.

---

Finally, no complete semantic model is available so far for intangible heritage or for the intangible component of the tangible one.

## 3. The way forward

In conclusion, there appears to be a need to improve the semantic apparatus of cultural heritage documentation. We suggest starting from and extending the CRM that is nowadays an internationally accepted standard. This will preserve the data interoperability across different documentation systems as they will all be compliant and based on the CRM. Even if two documentation systems introduce different specialization of CRM concepts, it is always possible to link elements across the two systems, availing, if necessary, of the more general CRM concept from which the two specializations are derived. This is what ARIADNE and 4CH are doing, the former by creating mappings between the original metadata schema of each dataset collection to be aggregated and a common ARIADNE one. 4CH is still in the design phase of its knowledge base and needs to incorporate provisions for heritage science and conservation data.

The need for an overarching documentation schema is motivated also by more general considerations. If different ontologies were used, interoperability might be put at risk unless mappings are provided. But the most important disadvantage would concern the methodological integration as advocated by the tandem metaphor: an overarching knowledge organization approach, common to the entire community of use, suitable for the different needs and adapted to the many existing research questions, is the first step (the bicycle, in the metaphor) for a real digital transformation of cultural heritage towards a common methodology. Such an approach must be suitable for all heritage documentation applications, from archaeological research to conservation. It needs to include the semantics required by a diverse research community. It must reconcile the different starting points and perspectives, from the documental one to the one organizing contributions of sciences like chemistry and physics, and moreover incorporate the architect's graphical approaches putting at the centre of documentation the shape, almost always interpreted as a 3D replica. We propose a unified methodology using the digital twin concept.

## 4. What is a digital twin

The concept of digital twin is not new. Digital twins have been applied in many industrial fields where the idea was born to test components, devices and, later on, to simulate the real behaviour of complex appliances in a digital way [16]. Then, digital twins made their way in machinery control applications, by using sensors surveying the behaviour of a device and sending an alert when an anomalous value is measured, or directly activating specific components to return to a normal condition. This kind of application required the use of simulation processes within the model, which were eventually incorporated in the concept of digital twin. Thus, for industry digital twins, the data component has a relatively simple schema, while the process part is more complex. The stress is on how the system behaves rather than on how the information about the system is structured. The European Commission has recently proposed to create a digital twin of the Earth [17] to evaluate complex environmental processes and their impact on the whole system, and to forecast the effects of mitigating measures against, for instance, global warming.

There are many digital twin definitions which put in evidence different characterizations according to the intended use. Two papers, [18] and [19], analyse many of them, relate them to different purposes and try to summarize them into a single overarching approach. None of the uses considered concerns cultural heritage. The 2020 paper by David Jones et al. titled "Characterising the Digital Twin: A systematic literature review" [18] mainly considers industrial applications, analysing in great details the features of such applications, with a short section dedicated to BIM. This thorough study may be useful in the future also for cultural heritage applications, as it analyses many aspects potentially to be considered when the interconnections between the dynamics of the real world of cultural heritage and those in the virtual world of digital twins will be modelled. Although the paper does not state a formal definition, its systematic approach collects and compares different naming conventions and may be instrumental to setting up a precise naming system for heritage applications.

Also the 2021 paper by E. VanDerHorn and M. Sankaran titled "Digital Twin: Generalization, characterization and implementation" [19] is based on a systematic literature analysis. The paper rightfully argues that "many of the definitions in the literature combine a definition with specific characterizations about Digital Twins that are unique to the use case(s) that they are describing", which creates confusion about the digital twin general definition. The overarching definition they propose is the following one:

> *[A digital twin is] a virtual representation of a physical system (and its associated environment and processes) that is updated through the exchange of information between the physical and virtual system.*

This shows that an information update is a substantial component of a digital twin system. However, we believe that the two phases of "virtual representation" and "update" are better analysed separately, in our case, where there is no straightforward way even for the first one.

The digital twin (DT) concept may also be described using the five-dimension description introduced by Qinglin Qi *et al.* [20]. Although still in a manufacturing perspective, their definition of a DT model may be useful also in the cultural heritage domain. They define a Digital Twin as made across five dimensions: Physical Entities (PE), Virtual Models (VM), Services (Ss), Data (DD), and Connections (CN). We would rename PE as *Real-world Entities* (RE), since this also encompasses the immaterial components of a real-world object. *Virtual Models* and *Services* include the behaviour of the Digital Twin according to the impulse of services: for example, how the Virtual Model of a building behaves when it receives the digital effects of an earthquake simulated by a Service. For cultural heritage, this kind of interaction is so far considered and processed outside the digital environment, making the virtual model still a descriptive one rather than a dynamic one. It is likely that in the future also simulation services will make their way into the heritage digital twin applications. *Connections* are the mutual interaction between any of the other four dimensions.

Digital twins are nowadays extensively used among others in mechanical engineering, architecture and especially in the building industry, where they belong to the BIM approach. A recent important project combining BIM and digital twins is the UK Gemini project, proposing to use digital twins nationwide for town planning [21].

Since the architecture domain has a close similarity with the cultural heritage (CH) one, such proximity has promoted the development of HBIM, i.e. Heritage BIM, which incorporates the BIM approach enriching it with additional classes pertaining to the heritage domain, but still within a flat data system. It seems that at present it is almost impossible to incorporate a much wider set of concepts – including some that are of a non-physical nature, the so-called intangible heritage – and relations among them. Recently, some authors [22, 23] suggested that an HBIM model is actually a Digital Twin and used it to develop applications in conservation of built heritage assets. For references to HBIM, see e.g. the bibliography published in [12] for archaeological applications and in general the survey [11].

In our opinion, the heritage domain requires much greater attention on how information is organised, an essential step before continuing to computational modelling that simulates real-world processes, the primary focus of most industrial digital twins. That is why we focus the present research on the data organisation, i.e. the semantics of digital twins, and for the time being set apart the digital simulation component so popular in the manufacturing applications. Thus, in this paper a *digital twin* is considered as *the complex of information about digital counterparts of real-world heritage objects, both material and immaterial ones*. We consider simulation as a distinct aspect, which includes processing heritage data in a computer environment, by creating virtual experiments on digital heritage objects in a virtual research environment to simulate the behaviour of the corresponding assets in the real world, or by availing of processes that get data from the real world via sensors, process such input in the virtual world using the features and data of the virtual models and, according to the results, update the virtual models and trigger real world actions via actuators. The data organisation within such a digital twin system has a relatively stable – but not immutable – design. Processes based on consuming the data to perform a simulation or other required operations may instead be created ad-hoc to address specific problems. Thus, the Heritage Digital Twin (HDT) discussed here does not (yet) consider dynamic interactions between Virtual Models activated by Services. It instead creates static Virtual Models based only on the knowledge derived by physical entities.

In sum, at present the Heritage Digital Twin is formed only by the knowledge about Real-world Entities, stored in digital format in the Virtual Model and the inclusion or extraction of such knowledge. Nevertheless, the information stored in the HDT is rather complex and needs a correct semantic structure, which is the purpose of the present paper. Services and Connections are not addressed so far. Indeed, introducing such dimensions would be useful for cultural heritage documentation and management, above all for its conservation. We plan to address these facets when the ontology is stable and has been reviewed by the community.

Our approach improves the current HBIM approach, strongly influenced by its CAD derivation and its architecture-civil engineering provenance. In HBIM, the Virtual Model consists of a (virtual) shape, almost always a 3D model, optionally divided into parts – e.g. the roof, the columns, the walls, etc. Each of them has some attributes such as colour, material, technical features, and so on. The corresponding semantic graph is a tree, having as root the graphical entity, typically a 3D model. With this approach, a set of information about a building would be worthless if no drawing is included, regardless of the richness of such information, while a 3D model would be worth consideration even if no information is included besides shape. This approach is equally present in the literature about CAD-based HBIM and in the other branch based on point clouds, where virtual models are called "augmented objects" and the respective non-graphical data often described as "annotations", a name that perhaps underlines their consideration as ancillary information compared to the graphical one. An interesting approach is proposed by CHER-Ob [24], which adds the need for additional information to the usual 3D annotation-based system. This approach is based on the concept of Cultural Heritage Entity (CHE) which corresponds to our Heritage Digital Twin. On the other hand, the main goal of CHER-Ob is more limited than ours, as it focuses on producing visual content for storytelling from available data, in what they call a *project*, i.e. a study on one or more CHEs addressing specific research questions. Information within a CHE is labelled according to the Getty Categories for the Description of Work of Art (CDWA) [25].

Differently from the 3D-based data models described in the current literature on heritage applications, we present here a semantic model, the Heritage Digital Twin ontology, in which there is no privileged class. The shape of an object is an important feature when it exists, but with no higher rank compared to others. For the above-mentioned reasons, we define the whole digital representation of the real-world CH 'object' as its digital twin, which consists of the aggregation of different components, among others its shape represented by a visual model. This approach as an extension of the CRM domain model also allows the creation of digital twins of immaterial 'objects' – having no shape by definition – i.e., in our case, intangible cultural heritage; of stories about heritage; and of people's relationships with tangible assets. Our proposed ontology incorporates the HBIM approach, which covers only a part of the information, and the 3D annotation model. A final consideration concerns the impact of time on digital twins. In industrial applications what matters is the present state, which can change but only the latest state is relevant. On the contrary, in cultural heritage applications information about past states may be relevant as the information of current ones. For this reason, we will timestamp all the information, i.e. consider a time span of validity. By default, this time span is "always", unless stated otherwise. At implementation, some simple service may automatically change this default assignment, for example always putting as beginning the construction date or putting as end the destruction date. Other cases concern documented changes, which may also be automatically dated. Time is defined according to ISO 8601-1 and ISO 8601-2. Note also that PeriodO[10] offers solutions to convert named periods (e.g. *The Renaissance*) into time intervals.

The approach presented here was initiated in a seminal paper [26] by some of the authors of the present one.

## 5. The Heritage Digital Twin: a discursive introduction

A necessary premise concerns the overall approach to the definition of the Heritage Digital Twin (HDT) ontology. The HDT ontology is based on the CIDOC CRM and its ecosystem, using whenever possible, the classes and properties of the CIDOC CRM and its extensions and defining new ones only to describe more specific concepts such as cultural entities, digital twins, 3D models, which do not find an exact match in the CIDOC CRM. In any case, as far as possible, we have always tried to derive the new entities from those of the CIDOC CRM in order to keep our model completely compatible, consistent and aligned with it. Introducing new classes maintains compatibility and interoperability when the new class is a (proper) subclass of an already existing one: if not possible at the subclass detail level newly introduced, interoperability will still be maintained at the superclass existing one.

---

[10] https://perio.do/en/

In the real world there are **Heritage Entities**, regarded as valued by a community – from the whole of humanity represented, for instance, by worldwide organizations as UNESCO, to smaller ones such as a group of believers. We do not enter into the debate of what is cultural heritage: whatever it is, the system is able to take it into account. A distinction is made between Tangible Entities and Intangible ones. A Heritage Entity may be composed from both tangible and intangible entities. A purely intangible heritage entity may be recognized as it has no tangible component.

The corresponding complex of (digital) information about a Heritage Entity is its **Heritage Digital Twin** (**HDT**). The term "complex" used above does not imply that the information must be stored in the same digital device: it can be distributed in different storage as long as they are accessible from the "main" system managing the HDT and possibly integrating on demand such a distributed knowledge. Thus, in such a main system information may consist in the actual data or in the data URI, resolved on demand when necessary.

This is the result of a holistic approach, incorporating, putting into order, and relating to it all the digital information pertaining to that Heritage Entity, and possibly to others as well. Thus, the Heritage Digital Twin is actually a network of relationships among data putting into evidence the connections between those data and a real-world Heritage Entity. There must obviously exist some item belonging to a Heritage Digital Twin, although it might conceptually be considered also when empty, a sort of placeholder for forthcoming digital information concerning some Heritage Entity. For example, a folder in a computer server containing photos, descriptions, e-books, and so on, of a particular monument is an embryonic heritage digital twin: not yet a full-fledged one because the relationships between each folder item and the heritage entity is not explained, it is just "the file (or a symlink to it) that stays in the folder named after the heritage entity".

The Heritage Digital Twin will then consist of pieces of its own information (e.g. the Twin's identifier) and of other digital information pertaining to it. Such information is organized as follows.

The main class is **Heritage Entity**, comprising tangible and intangible entities of the real-world regarded as valuable because of their contribution to society, knowledge and/or culture. The tangible and intangible aspects of the *same* Heritage Entity are recorded as **Tangible Aspect** and **Intangible Aspect**. While all Heritage Entities have an intangible aspect, some may not have the tangible one.

As already mentioned, the complex of the information concerning a Heritage Entity forms its Heritage Digital Twin. The latter includes **Digital Representations**, i.e. digital representations of a Heritage Entity such as text documents or visual ones, i.e. a photo, a video, a 3D model, each one with its own class. It also includes **Stories** about the Heritage Entity, i.e. narratives, modelled according to the NOnt ontology [27, 28], which is based on narratology, a formal way to describe narratives. It distinguishes the *fabula*, i.e. the thematic content of a narrative, and the *narration* (also called *syuzhet*), the chronological structure of the events within the narrative.

The resulting HDT ontology is compatible with the CRM and is also extensible. For example, further subdivisions of text documents might be introduced if necessary, distinguishing e.g. among historical documents, scientific documents and so on. 3D models may be further characterized according to type and have their own metadata and paradata as required. Here we introduce only the most general classes, and are planning to present such extensions in a forthcoming paper.

## 6. Technical description of the HDT ontology: introduction

### 6.1 General note on classes and properties

As already mentioned, the HDT ontology is based on the CIDOC CRM and its ecosystem, directly using the classes and properties of the CIDOC CRM and its extensions for entities and properties having identical meaning and conceptual scope also in our domain, such as places, agents, physical objects. In this case the class or property name is preceded by the namespace prefix, which identifies the relevant ontology.

### 6.2 Ontological models used in HDT ontology

The following Table 1 shows the ontological models of the CIDOC CRM ecosystem used to build the HDT ontology and the namespaces (prefixes) used to indicate their classes and properties throughout this document[11].

Table 1. Ontological models used in HDT ontology definitions

| Model | Version | Name space prefix | Description | Classes prefix | Properties prefix |
|-------|---------|-------------------|-------------|----------------|-------------------|
| HDT | 1.0 | hdt | The ontology described in the present paper (i.e. new classes and properties) | HC | HP |
| CIDOC CRM | 6.2.1 | crm | A formal ontology for modelling Cultural Heritage information | E | P |
| CRMsci | 1.2.6 | crmsci | The scientific observation model | S | O |
| CRMdig | 3.2 | crmdig | Model for provenance metadata | D | L |
| CRMpe | 3.1.2 | crmpe | The PARTHENOS Entities model | PE | PP |
| CRMinf | 0.10.1 | crminf | An Extension of CIDOC-CRM to support argumentation | I | J |
| CRMba | 1.4 | crmba | An extension of CIDOC CRM to support buildings archaeology documentation | B | BP |
| FRBRoo | 2.4 | frbr | Functional requirements for bibliographic records | F | R |
| NOnt | 1.0 | nont | The MINGEI Narrative Ontology | | |

**6.3 Events**

To model cultural events, traditions and practices, typical of the intangible heritage, we have defined the HC4 Intangible Aspect class, which is declared as a subclass of crm:E89 Propositional Object and is referred to by an HC3 Tangible Aspect class by the property HP5 has intangible aspect. While all tangible heritage has always an intangible aspect, intangible heritage does not necessarily have a tangible aspect.

Instances of HC4 Intangible Aspect describe generic (template) events such as the Palio di Siena and not the individual occurrences of the Palio. The actual individual occurrences are instances of crm:E5 Event, and property crm:P129 is about (is subject of) may be used to link an individual Event (E5) to the generic description of the cultural heritage event (HC4), such as the Palio di Siena race of the present year (E5) which is about (P129) the Palio tradition (HC4). Nevertheless, it is important to define a new property to specify this special link between the HC4 Intangible Aspect and its punctual manifestations. This property is HP6 has manifestation event (event is manifestation of). Notice that, with this minimal requirement model, recurrent event series as specified in [29] may be computed, instantiated and associated to an HC4 instance if needed. The following is an example:

The Palio di Siena (HC1 Heritage Entity), is a horse race (HC4 Intangible Aspect) that is held twice each year, on 2 July and 16 August, in Siena, Italy (crm:E53 Place). Ten horses and riders (crm:E39 Actor), represent ten of the seventeen *contrade* (crm:E74 Group), or city wards.

The historical horse race Palio di Siena (HC4 Intangible Aspect) was held again (crm:E5 Event) on 17/8/2022 (crm:E52 Time-Span) after a two-year pause because of the COVID-19 pandemic. The winner was jockey Giovanni Atzeni (crm:E39 Actor).

For events and activities that are not strictly "cultural" and therefore fall outside the immediate scope of HC4 Intangible Entity, but which in any case concern, affect and remain somehow connected to the various cultural entities, the crm:E5 Event class can be used instead. These are activities like conservation, restoration, reconstruction, and natural events like earthquakes and floods. The type of events for which instances of crm:E5 are used can be specified by means of the crm:P2 which has type property of the CIDOC CRM. Open vocabularies, containing the most common descriptions of these cultural heritage activities, can be defined and released within the ontology to be used in combination with the above-mentioned crm:P2 property for a more standardized and complete description.

## 6.4 Conditions and states

An important information about heritage assets is their condition state. In the CRM this is documented using E3 Condition State followed by P2 has type E55 Type, choosing the latter in a vocabulary of possible states.
If this way of assessing the state of an asset seems too generic, a more precise solution is offered by E14 Condition Assessment. This is the activity dedicated to the evaluation of the condition, and its outcome is a report about the condition. For example, a paper titled "Three-Dimensional Creep Analyses of The Leaning Tower of Pisa" on the condition of the Pisa leaning tower was published in 1997 by Dryden and Wilson. This fact can be expressed as follows:
The "Analysis on Pisa Tower" (E14 Condition Assessment) in 1997 was carried out by (P14 carried out by) "Dryden & Wilson" (E39 Actor) in (P4 has time span) 1997 (E52 Time-Span). It is documented (P70 documents) in the document "Three-Dimensional Creep Analyses of The Leaning Tower of Pisa" (E31 Document).

## 6.5 Stories and Storytellings

In our model, stories are considered as accounts of facts about a certain Heritage Entity, including (but not limited to) descriptions based on documents and on their interpretation. Stories are an integral aspect of the framework of a Heritage Entity (and thus, of its Digital Twin) since they contribute to the construction of its intangible part as they may make it more understandable, interesting, and attractive for the public. A story can be seen as the core of a series of facts and how they happened, as for example the story of the Knossos Palace and its discovery by Arthur Evans or the story of Falconry over the centuries. From a conceptual point of view, a story is equivalent to the concept of *narrative*, i.e., "a story as it happened in reality or in fiction", as defined in the Narrative Ontology (NOnt) of the MINGEI project[12]. Since this definition is perfectly suited to our concept of story, and furthermore, given that Narrative in the Narrative Ontology is a subclass of E73 Information Object of CIDOC CRM, which provides it an additional level of formal compatibility, this class will be used in our model as it is.

A storytelling, instead, is considered in our model as the way in which the facts composing a story are actually narrated, presented and disseminated. Storytellings, in fact, comprise social and cultural activities of telling, writing and disseminating stories, for the purpose of education, cultural preservation or entertainment, both in oral form and by means of simple or sophisticated techniques aimed at making the narration of a story effective. An example of storytelling is how the history of Falconry is narrated in "De Arte Venandi Cum Avibus" treatise by the Holy Roman Emperor Frederick II. From a conceptual point of view, a result of storytelling can be seen as equivalent to the concept of narration, which in the Narrative Ontology "represents the narration of a narrative, i.e. an individual work that tells the events of the narrative through some form of media (text, video, audio, etc.)". As stated in [27, 28] there can be many narrations of the same

---

story, focusing on different aspects of the *fabula*, or presenting events in a different order. In the Mingei Narrative Ontology, Narration is a subclass of F14 Individual Work of the FRBRoo Ontology, another extension of CIDOC CRM. Since also in this case, as for the story, we have verified the perfect conceptual and formal overlap of this class with our idea of storytelling, the Narration class will also be used in our model as it is.

## 7. Scope notes of Classes and Properties

### 7.1 Classes

### HC1 Heritage Entity

Subclass of:     crm:E77 Persistent Item

Superclass of:

Scope Note:      This class comprises tangible and intangible entities of the real-world regarded as valuable because of their contribution to society, knowledge and/or culture. Instances of HC1 Heritage Entity may refer to real assets of any nature: physical, both movable and immovable, immaterial, or born digital. They can also refer to cultural events, traditions and practices, typical of the intangible heritage, and can be used to describe their features and their extent in space and time. In the case of events, we can create instances of event types. An instance of HC1 can be considered as the entry point for inferring the content of its corresponding HC2 instance, even if by using crm:P148 has component property it can also be used for denoting HC2's components.

Examples:        the Knossos Palace, part of the Knossos WH archaeological site
the Pafos Gate in Nicosia
the "Palio di Siena"
the Florence Historical Centre, a WH Site
the Stonehenge Complex, a WH site
the Bauhaus style.

Properties:      HP1 has digital twin (is digital twin of): HC2 Heritage Digital Twin
HP2 has story (is story about): nont:narrative
crm:P70 is documented in (documents): HC6 Digital Heritage Document
HP9 has visual representation (is visual representation of): HC7 Digital Visual Object
crmdig:L1 was digitized by (digitized) D2 Digitization Process
crmdig:L11 had output (was output of): HC8 3D Model[13]

### HC2 Heritage Digital Twin

Subclass of:     crm:E89 Propositional Object

Scope Note:      The class consists of the information available in a given system and pertaining to an HC1 Heritage Entity.  Every instance of HC1 Heritage Entity is linked to one instance of HC2 Heritage Digital Twin, which provides an archive of the documented history of the corresponding HC1 Heritage Entity. It includes digital representations of that Heritage Entity (e.g. 3D models, images, videos), textual descriptions (e.g. digital documents, narrations or stories), information of the effects on the related HC1 Heritage Entity of events that influenced or/and are related in any way to its state of (e.g. earthquakes, floods etc.) and of activities (e.g. restorations, conservations etc.) carried out on it.

---

[13] This means that an HC1 instance is related to an HC8 instance via the property path crmdig:L1/crmdig:L11.

Examples:          the HDT of Pisa Leaning Tower
                   the HDT of the Neptune Fountain in Bologna
                   the HDT of Knossos Palace
                   the HDT of the Pafos Gate in Nicosia
                   the HDT of the "Palio di Siena"
                   the HDT of the Florence Historical Centre

Properties:        HP3 is digital twin component of (has digital twin component): HC2 Heritage Digital Twin

## HC3 Tangible Aspect

Subclass of:       HC1 Heritage Entity
                   crm:E18 Physical Thing

Scope Note:        This class comprises tangible, material entities of the real-world, both movable (e.g.
                   archaeological, artistic and cultural objects) and immovable (e.g., built heritage like
                   monuments, buildings, cities and other complexes), regarded as valuable because of their
                   contribution to society, knowledge and/or culture. The "tangible" term in the name of this
                   class does not exclude that its instances also possess an intangible aspect, which is specified
                   through the HP5 has intangible aspect property.

Examples:          the Neptune Fountain in Bologna (Italy)
                   the Pisa Leaning Tower, a UNESCO World Heritage (WH) Site
                   the Nike of Samothrace of the Louvre Museum in Paris (France)

Properties:        HP5 has intangible aspect (is intangible aspect of): HC4 Intangible Aspect
                   HP7 is manifestation of (is manifested by): HC4 Intangible Aspect

## HC4 Intangible Aspect

Subclass of:       HC1 Heritage Entity
                   crm:E89 Propositional Object.

Scope Note:        This class comprises cultural events, traditions and practices having particular social,
                   historical and cultural significance, including practices and expressions, memories and oral
                   traditions about events, things, people.

Examples:          the Mediterranean diet
                   Falconry
                   the Rebetiko music tradition
                   the "Palio di Siena"

Properties:        HP6 has manifestation event (event is manifestation of): crm:E5 Event

## HC5 Digital Representation

Subclass of:       crmdig:D1 Digital Object

Scope Note:        This class comprises the digital virtual representations of an HC1 Heritage Entity such as e-
                   texts, images, audio or video items, 3D models, etc., that are documented as single units.

Examples:          the digital version of Vasari's "Vite"

the video https://www.youtube.com/watch?v=P1Uv4Zf5xKk

the Pafos Gate laser scanning 3D model

## HC6 Digital Heritage Document

Subclass of:    HC5 Digital Representation

Scope Note:    This class comprises pieces or collections of digital, non-visual documents, either born-digital or digitised from physical, real-world ones, typically containing textual or numerical information regarding an HC1 Heritage Entity and intended to become part of the related HC2 Heritage Digital Twin. Documentation of this kind may include scientific data, research results and interpretation, as well as historical and cultural information, including textual descriptions related to the nature, conditions, positioning and to the whole set of events in which the cultural entity has been involved and the actors who have participated in them.

## HC7 Digital Visual Object

Subclass of:    HC5 Digital Representation

Scope Note:    This class comprises digital visual objects, such as photos and videos, but also special imagery such as X-ray images, spectra of chemical and physical analyses, and so on, intended to become part of the HC2 Heritage Digital Twin of an HC1 Heritage Entity. Digital documentation of this kind can be born digital or digitised from physical objects (such as paper photographs, drawings and so on). Particularly relevant digital visual objects are also Virtual Reality (VR) and Augmented Reality (AR) models, other types of visual digital artefacts pertaining to a HC1 Heritage Entity. Both VR and AR models rely on 3D models of the related heritage entity, but may add or remove parts of it, or require further digital input as in AR, so they should be catalogued separately from 3D models.

Examples:    The Europeana digital version of the paper picture of the Pisa Leaning Tower taken by Paolo Monti in 1960 (https://www.europeana.eu/it/item/9200369/webclient_DeliveryManager_pid_6363979_custom_att_2_simple_viewer).

## HC8 3D Model

Subclass of:    HC5 Digital Representation

Scope Note:    This class is used for rendering in detail the 3D model of HC1 Heritage Entity and intended as a particular crmdig:D1 Digital Object having its definite identity and resulting from operations such as digitization, acquisition, processing and other actions typical of the three-dimensional modelling world (e.g., 3D scanning, wireframe modelling and so on). The particular features of a 3D model (e.g., its type, format, resolution, size, etc.) and its relationships with the series of activities carried out for its creation and manipulation are modelled through the properties inherited from its superclass HC5, which in turn inherits from crmdig:D1 Digital Object, and through the other classes and properties of CRMdig.

Examples:    The 3D model of the Neptune Fountain produced by ISTI-CNR (Pisa, Italy) as part of the documentation used for the restoration of the Neptune Fountain in Bologna (Italy). https://www.cnr.it/en/focus/074-43/3d-supported-restoration-the-neptune-fountain-in-bologna.

## 7.2 Properties

### HP1 has digital twin (is digital twin of)

Domain:        HC1 Heritage Entity
Range:         HC2 Heritage Digital Twin

Scope Note:    This property links an instance of HC1 Heritage Entity with an instance of its related HC2 Heritage Digital Twin in a given system.

Examples:     The Pafos Gate in Nicosia, Cyprus (HC1) has digital twin (HP1) the Pafos Gate digital twin (HC2) created by Cyprus Institute.

### HP2 has story (is story about)

Domain:        HC1 Heritage Entity
Range:         nont:Narrative

Scope Note:    This property links an instance of HC1 Heritage Entity with an instance of a nont:Narrative that refers to it.

Examples:     Falconry has story (HP2) the history of Falconry over the centuries.

### HP3 is digital twin component of (has digital twin component)

Domain:        HC2 Heritage Digital Twin
Range:         HC2 Heritage Digital Twin

Scope Note:    This property associates an instance of HC2 Heritage Digital Twin with another HC2 of which is component. The term 'component' here is not limited to physical or geographical relationships (see examples), but encompasses any kind of main-associated relationship.

Examples:     The HC2 Digital Twin of Pafos Gate in Nicosia (Cyprus) HP3 is a digital twin component of the HC2 Digital Twin of Nicosias' City Walls.
The HC2 Digital Twin of the "Cento Camini" Medici Villa in Artimino (Florence) HP3 is a digital twin component of the HC2 Digital Twin of the UNESCO WHS Medici Villas in Tuscany.
The HC2 Digital Twin of Vichy is a HP3 digital twin component of the HC2 Digital Twin of the UNESCO WHS The Great Spa Towns of Europe
The HC2 Digital Twin of the "Basilica of San Salvatore in Spoleto, Italy" is a HP3 digital twin component of the HC2 Digital Twin of "Spoleto", which is a HP3 digital twin component of the UNESCO WHS "Longobards in Italy. Places of Power"

### HP4 narrates (is narrated through)

Domain:        nont:Narration
Range:         nont:Narrative

Scope Note:    This property links an instance of nont:Narration with an instance of a nont:Narrative which has this narration. It is similar to the nont:hasNarration property, but is not a subproperty of crm:P148 has component.

Examples:     The *De Arte Venandi Cum Avibus*" treatise by the Holy Roman Emperor Frederick II narrates (HP4) the history of Falconry.

**HP5 has intangible aspect (is intangible aspect of)**

Domain:        HC3 Tangible Aspect
Range:         HC4 Intangible Aspect

Scope Note:    This property associates an instance of HC3 Tangible Aspect with its intangible aspects (HC4), i.e. the cultural, social and historical value it incorporates.

Examples:      The "Theotokos of Vladimir" (HC3) icon HP5 has intangible aspect the secular veneration that is addressed to it (HC4).
The UNESCO WHS site "Routes of Santiago de Compostela" (HC3) has intangible aspect (HP5) pilgrimage to Santiago (HC4).

**HP6 has manifestation event (event is manifestation of)**

Domain:          HC4 Intangible Aspect
Range:           crm:E5 Event
SubPropertyOf:   crm:P129 is about (is subject of)

Scope Note:    This property associates an instance of HC4 Intangible Aspect with the instances of the crm:E5 Event (or of the unique and specific crm:E5 Event) through which the intangible entity manifests itself in the physical world.

Examples:      The Palio di Siena (HC4) has manifestation event (HP6) the historical horse race that was held in Siena on 17/8/2022 (E5)

**HP7 is manifestation of (is manifested by)**

Domain:        HC3 Tangible Aspect
Range:         HC4 Intangible Aspect

Scope Note:    This property associates instances of HC3 Tangible Aspect with the HC4 Intangible Aspect of which they are the manifestation in the physical world.

Examples:      The set of devotional graffiti engraved on the walls of the Church of the Holy Sepulchre in Jerusalem (HC3) is the manifestation of (HP7) the pilgrimage of which the church is the final destination (HC4).

**HP8 is narrated in document (document used for narration)**

Domain:        nont:Narration
Range:         crm:E31 Document

Scope Note:    This property associates an instance of nont:Narration with instances of E31 Document used to implement it.

Examples:      The "De Arte Venandi Cum Avibus" treatise by the Holy Roman Emperor Frederick II (nont:Narration) is narrated in document (HP8) the "MS. Lat. 419" manuscript , now in the library of the University of Bologna E31.

**HP9 has visual representation (is visual representation of)**

Domain:        HC1 Heritage Entity
Range:         HC7 Digital Visual Object

Scope Note: This property associates an instance of HC1 Heritage Entity with instances of HC7 Digital Visual Object in which it is represented.

Examples: The Pisa Leaning Tower (HC1) has visually representation (HP9) the Europeana digital version of the paper picture of the Pisa Leaning Tower taken by Paolo Monti in 1960 (https://www.europeana.eu/it/item/9200369/webclient_DeliveryManager_pid_6363979_c ustom_att_2_simple_viewer) HC7

**HP10 tells about (is told by)**

Domain: nont:Narrative
Range: crm:E5 Event

Scope Note: This property is intended to identify the specific events (E5) to which a nont:Narrative relates.

Examples: The history of Falconry (nont:Narrative) tells about (HP10) the writing of "De Arte Venandi Cum Avibus" treatise by the Holy Roman Emperor Frederick II (E5).

## 8. Use cases

In this section we develop two complete examples of HDT. In the first one, the focus is on art history and on scientific analyses carried out on a painting. The second one describes the monastery of St. John Lampadistis in Kalopanayotis, Cyprus, including the description of intangible aspects.

### 8.1 The analyses on the portrait of Caterina Cornaro from the Leventis museum of Nicosia

### 8.1.1 History of the painting and results of the analyses carried out on it

The painting was acquired by a private collector (Michael Zeippekis) who donated it to the Leventis museum. It is a 19$^{th}$ century portrait of Caterina Cornaro, the last queen of Cyprus [30]. Visual investigations hinted at the existence of an underlayer of paint. Further heritage science investigations (x-ray imaging, digital microscopy, multi-spectral imaging, XRF) confirmed the existence of an earlier, 16$^{th}$ century painting [31]. Tobias Lange, a restorer from Dresden, Germany, removed the 19$^{th}$ century painting and exposed the 16$^{th}$ century one (Figure 1). Many details changed: the crown, turban, breast cover and right shoulder size and orientation.

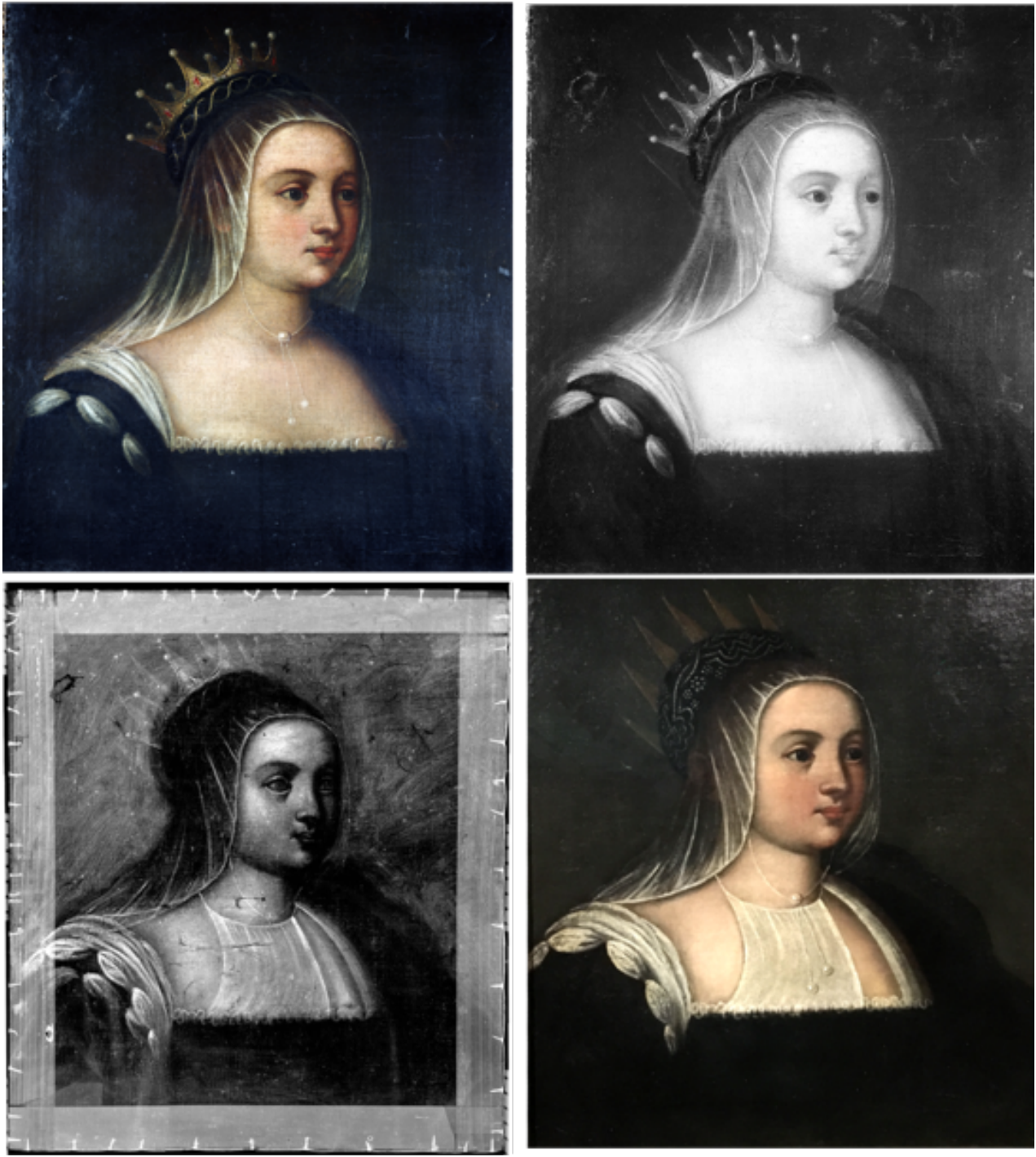

Figure 1. (a; b; c; d: clockwise from top left): The painting's development: a – as appeared before latest restorations; b – infra-red imaging reveals an underlayer paint with a modified pictorial composition; c – X-Ray imaging displaying in more details these differences: shoulder in a different position, a cloth covering the chest, the veil position shifted, a pointed, un-decorated crown, the turban broader and richer, the ear more to the left; d – the painting as of today, after restorations that removed the upper pictorial layer (added in the 19<sup>th</sup> century) and highlighting the 16<sup>th</sup> century original.

Art historical investigations indicate a lost Titian portrait, which was the source of inspiration for several copies, made by different artists during the 16<sup>th</sup> century and which circulated in Europe. Common to all is the shape of the pointed crown, the queen depicted as a mourning widow with characteristic clothes and a veil.

A close examination of the painting revealed some details implying relevant information about the painting. Boldface indicates the parts developed in the semantic example.

Table 2. Results deduced from observation of the painting and from literary references, in boldface those modelled with HDT in the example

| Observation | Inference |
| --- | --- |
| **Painting canvas nailed on stretches** | **Painting cut from a larger one** |
| No signatures | A 19th century painting overpaint on a 16th century one, unknown artists |
| Old adhesive on the retro - GK9654 | GK = Royal Prussian Cataloguing system |
| Old adhesive on the retro in German – mentioning the collectors | Edward Solly was an English collector focusing on Renaissance paintings – he donated his collection in the 19th century to the Royal Prussian house. |
| Observations on a cross-section sample from the stretcher with a transmitted light microscope (technical details mentioned) | Wood structure indicates poplar (Populus), widely used in S. Europe, but not N. Europe, to make stretchers, indicating that it was probably made in Italy. |
| References in various books and collection inventories (Vasari's Lives, Colbertaldo's Storia di Caterina Cornaro) on paintings | Portraits of the queen by several Venetian artists, such as Titian, Giorgione, Tintoretto, Paolo Veronese, etc., done from imagination. |
| **Carlo Ridolfi writes in his Le Meraviglie dell' Arte on Titian's work "… with the same way the Queen Caterina Cornaro is depicted in widow's dress that lets the whiteness of the skin rise through the black…"** | **Indication of a now-lost painting of Titian with an iconography of "Caterina Cornaro as a widow".** |
| Marin Sanuto – chronicler and member of the Major Council of Venice mentions in his diary that upon her arrival in Venice, the queen wore "a black velvet dress, a veil on her head and Cypriote jewellery". | Hanging pearls as "Cypriote jewellery", the black dress, the almost transparent veil and the chest cover became a fashion of depicting Cypriot nobility during the 16th century paintings in Cyprus. |
| Use of gesso on a tabby waved canvas, brown ochre (imprimatura) on the preparation layer, paint layer applied with fine brush strokes. | Techniques of a Venetian Renaissance painter. |
| Canvas of thin linen fabric in simple tabby weaves. | Produced in the Low Countries and widely used in Venice during the 16th century. |
| Thin gesso layer on the canvas. | Fills the depressions in the weave. |
| Brown underpaint layer on top of the gesso. | Typical of 16th century paintings. |
| No underdrawing beneath the painted surface following IRR and X-ray analysis. | Outlining of forms with a brush and fluid black was common to Titian's practice as evidenced on his autographed paintings. |
| No outlining of the canvas for a duplicating process, which would serve to duplicate the portrait to another surface. | This suggests a creation after a prototype, perhaps a repetition, worked with a free hand. |
| Later overpaints in various areas of the painting, such as re-positioning of the ear, change in the shape of the crown and turban, shoulders and the covering of the chest with a white fabric. | Efforts to bring the composition to the fashion of the 19th century, as well as to centralize better the figure, since it was initially cut from a larger painting. |
| Overall composition of the 16th century painting is of high quality. | As described by C. Ridolfi, this painting probably coincides with the times of Titian, but unclear by whom and is among the earliest among the queen's portraits. |

Table 3. List of Heritage Science experiments.

| Instruments and environmental conditions | | Reason |
|---|---|---|
| Imaging | Ricoh WG-30 digital camera | Delineation of composition details on pictographic layers. |
| | Hirox KH-8700 digital microscope with a magnification range of 5×–2500× with dual illumination revolver zoom lenses. | |
| Spectroscopy | XRF - ARTAX-200 μ-XRF Bruker, Mo X-ray tube operated at 50 kV and 700 μA, CCD camera with sample illumination and laser spot, a silicon drift detector with a resolution of <150 eV and a 0.65 mm collimator. Acquisition time = 90 s. Measurements taken in air atmosphere and with a Mo filter. The energy to channel calibration was done with a bronze standard, using the Cu- and Sn-Kα lines; the Mn-Kα line of a manganese standard for the full width at half maximum calibration. Spectra acquisition and evaluation were done with Spectra 5.3 software. | Identification of pigments used for all pictorial and preparation layers. |
| SEM | SEM-EDS - TM3000 Hitachi, with a Backscattered Electron (BSE) detector coupled with EDS for elemental analysis. The system was operated at 15 kV at a working distance of 8.5 mm; images were taken at magnifications of 15× to 30,000×. | Painting stratigraphy |

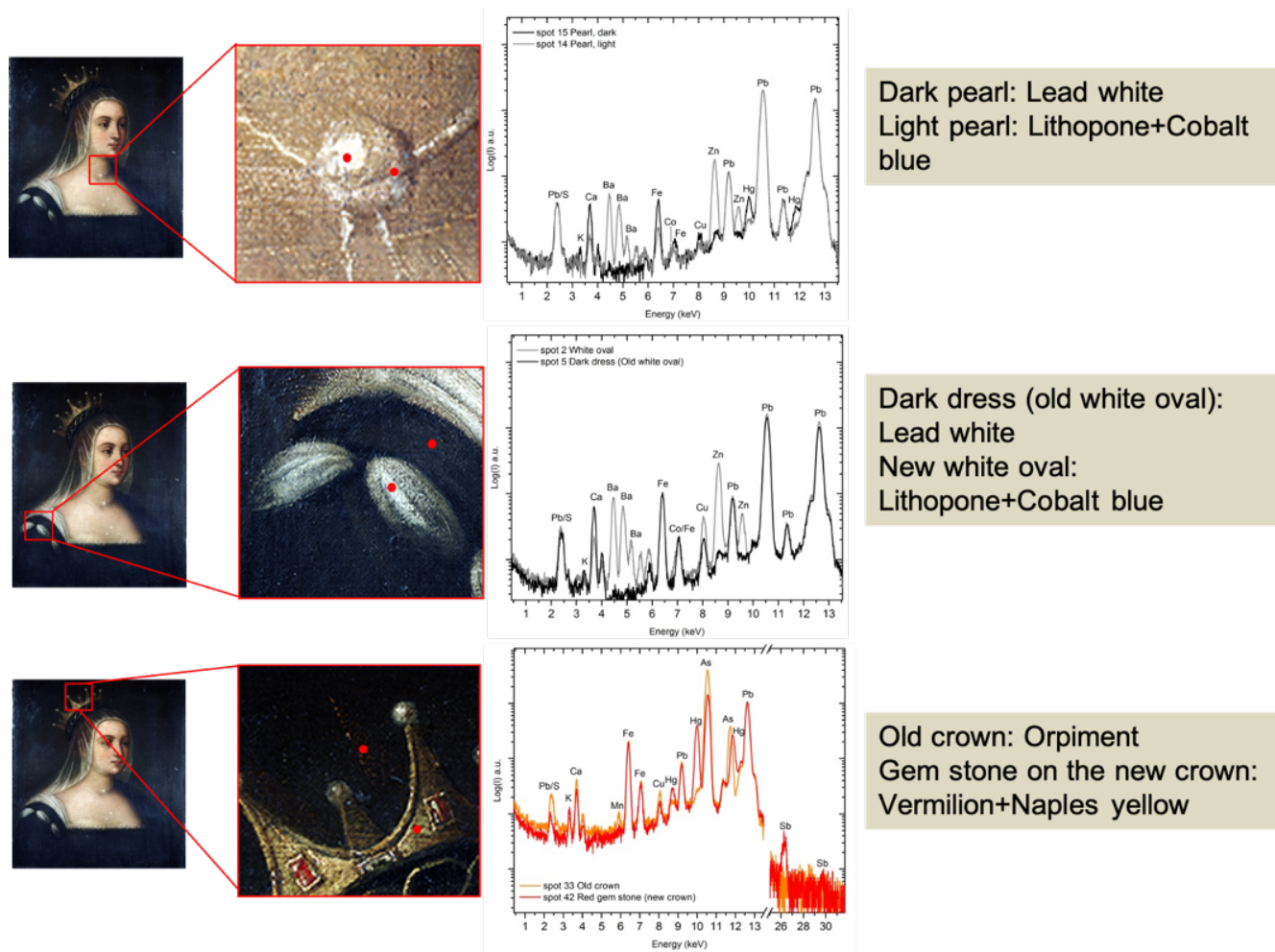

Figure 2. X-Ray fluorescence analyses of the various spots on the painting and suggested pigments used.

Table 4. Results of the analyses and conclusions based on them.

| | Observation | Inference |
|---|---|---|
| EDS | Intense Ca and S lines | Preparation layer of canvas– a thick white layer of calcium sulphate, either anhydrite ($CaSO_4$) or gypsum ($CaSO_4 \cdot 2H_2O$). |
| EDS | Pb, Fe, Ca, Mg, Na, Al and P. | Brown preparation layer called *imprimatura*, characteristic of Renaissance paintings consisting of a mixture of lead white, earth pigments (ochre), ivory black and dolomite (?). |
| | A thin organic layer. | Residues of varnish from the original painting. |
| | Two uppermost transparent layers, recognizable under UV light microscopy. | Layers of varnish of the overpainting. |
| | Staining tests on the painting cross- sections detected a proteinaceous material in a paint layer of the overpainting as well as in the Ca-containing white ground layer. | Presence of a glue binder. |
| XRR | Bright areas. | X-ray opaque pigments composed of heavy elements. |
| XRR | Dark areas. | X-ray pigments composed of light elements. |
| XRF | **Basic lead carbonate, intense Pb lines** | **Lead white ($2PbCO_3 \cdot Pb(OH)_2$), characteristic of Renaissance paintings.** |
| XRF | Intense Ba and Zn lines. | Lithopone (barium sulfate-zinc sulfide, $BaSO_4 \cdot ZnS$). Pigment used in the second half of the 19th century (thus dating the overpaint). |

Table 5. Colour Analysis with XRF

| | | Original paint | Overpaint |
|---|---|---|---|
| White | | Lead white ($2PbCO_3 \cdot Pb(OH)_2$) | A mixture of Lead white with Lithopone ($BaSO_4 \cdot ZnS$) |
| Carnation | | A mixture of Lead white, Vermilion (HgS), Ochre (iron yellow, brown) and Copper green. | |
| Yellow | | Mainly Orpiment ($As_2S_3$). Small intensity of Sn lines might be related to small additions of a yellow tin pigment, such as Lead-tin yellow ($Pb_2SnO_4$). Small intensity of Ag lines might be related to admixtures of silver in natural orpiment. | Mainly Lead yellow (Litharge PbO) |
| Red | none | Vermilion for the lighter and paler red, Vermilion with an antimony pigment such as Antimony red ($2Sb_2S_3 \cdot Sb_2O_3$, superposition or mixture) used to highlight the borders of the gemstones which appear brighter and darker. | |
| Green | | | Copper green |
| Black | | Iron-Manganese black (Umber) and Copper (green) pigments | |

Each statement from the tables above is supported by available datasets.

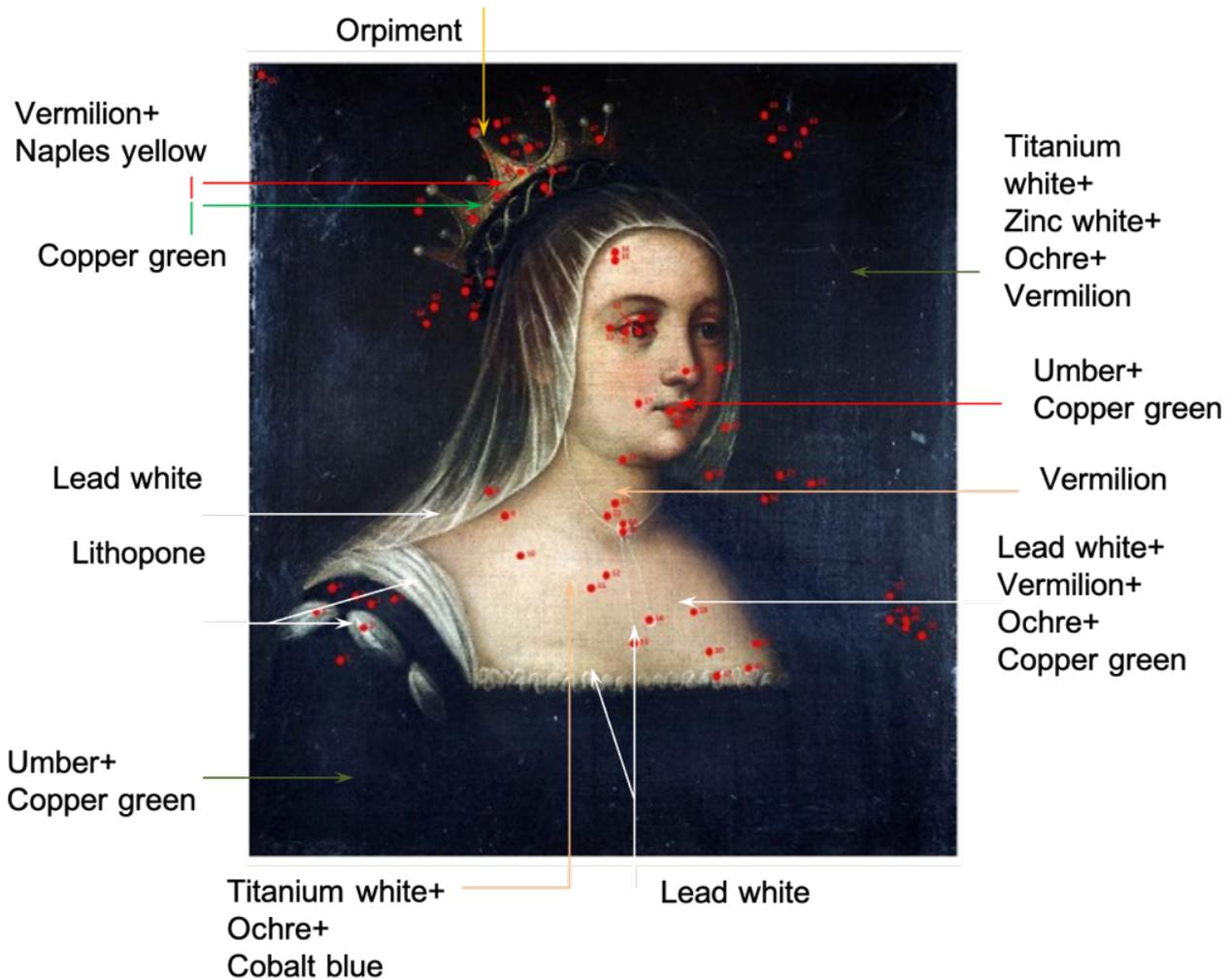

Figure 3. Distribution of pigments on the painting, demonstrating a mixture of original 16th century ones and later, 19th and 20th century pigments

### 8.1.2 Modelling Art History information

As a preliminary test, we modelled two of the results proposed above (highlighted in boldface). The first concerns a direct observation of the condition of the painting, the second concerns information derived from a book in which the painting is probably mentioned.

In particular, in the first case, some physical peculiarities are observed (HC1 → O8 → S20) on the painting canvas (O9 → S9) which allow researchers to make hypotheses (J2 → I2) about past conditions and provenance of the object.

In the second case, from Carlo Ridolfi's book (E31), in which the object is mentioned (HC3 → P70), we read that Titian painted a portrait of Caterina Cornaro, now lost (J7 → I7). This leads to the hypothesis (J2 → I2) that this portrait could be the lost one painted by Titian.

Finally, the two I2 Belief(s) resulting from the above observations (together with the others from Heritage Science described below, and others deriving from further investigations and analyses not modelled here) can then be used as a premise (J1) to support further inferences (I5), such as the attribution of the painting to Titian (I2).

The full model is shown in Figure 4.

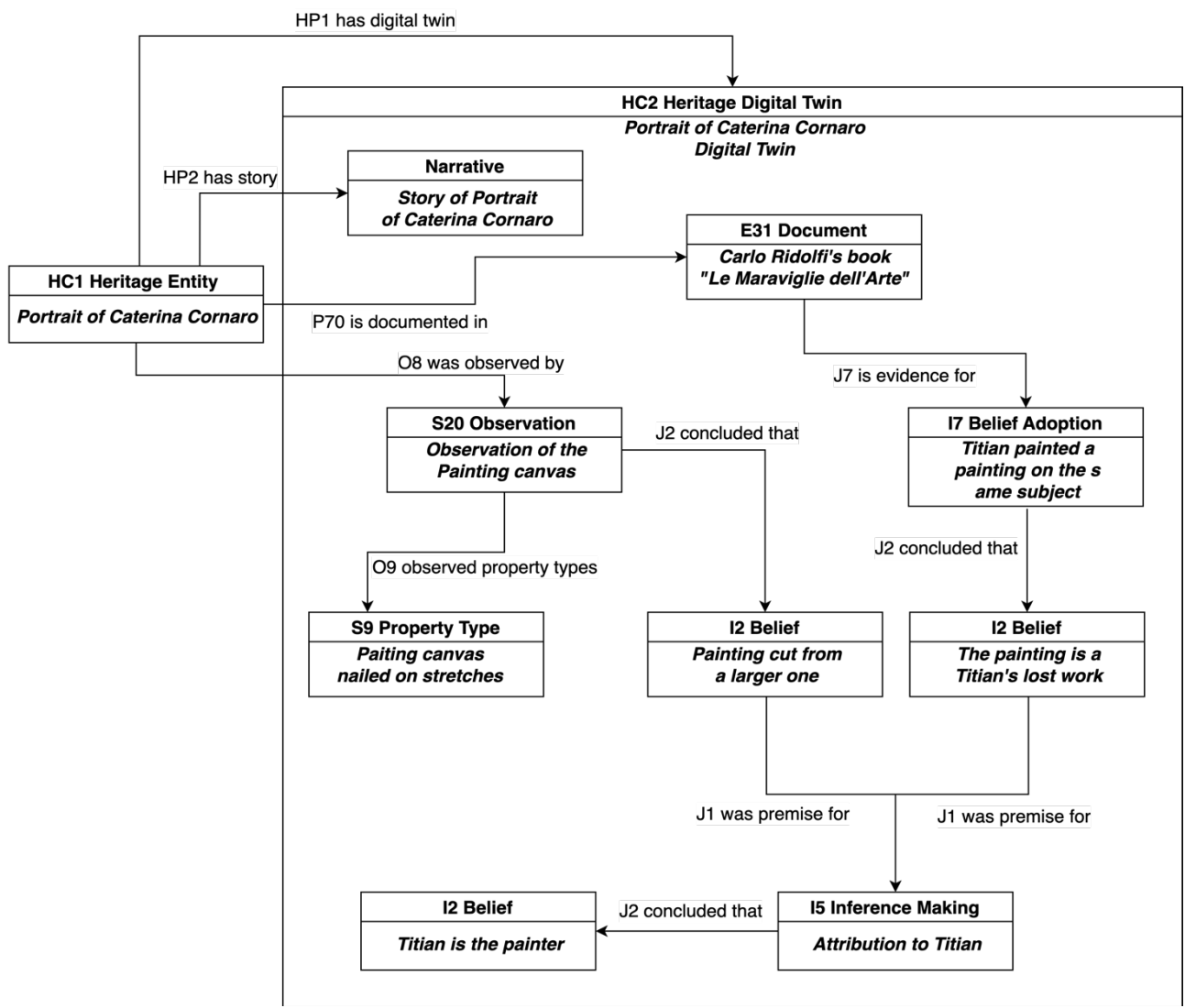

Figure 4. Modelling Art History research activities

### 8.1.3 Modelling Heritage Science results

For the Heritage Science test, we chose spectroscopy and x-ray imagery as representative examples of scientific investigations (also highlighted in boldface in this document, see above). In this case, the information derives from a spectroscopic examination of the painting (HC1 → P39 → D11), intended to identify specific pigments (P17 → E7), that was performed using a specific device (L12 → D8) equipped with additional physical accessories (P16 → E19), operated by specific software (L23 → D14) and using a series of settings and environmental parameters (L10 → D9).

The spectroscopic analysis produced a series of results (L20 → D9) which, once evaluated (O16 → S6 → O11 → S9), led the experts to some conclusions (J2) regarding the chemical elements used by the artist as characteristic of a certain historical period (I2), in this case the "Renaissance".

The painting was also analysed by means of x-ray imaging techniques (L1 → D2) which created an x-ray image (HC7) of it, through which bright areas present on the painting are observed (O8 → S20 → S9) and are associated (J2) by the experts with the use of pigments composed of heavy elements (I2).

Also in this case, as for Art History, the final scientific results can constitute the bases (J1) for the attribution of the painting to a certain author (I5 → J2 → I2), in this case to Titian. The full model is shown in Figure 5.

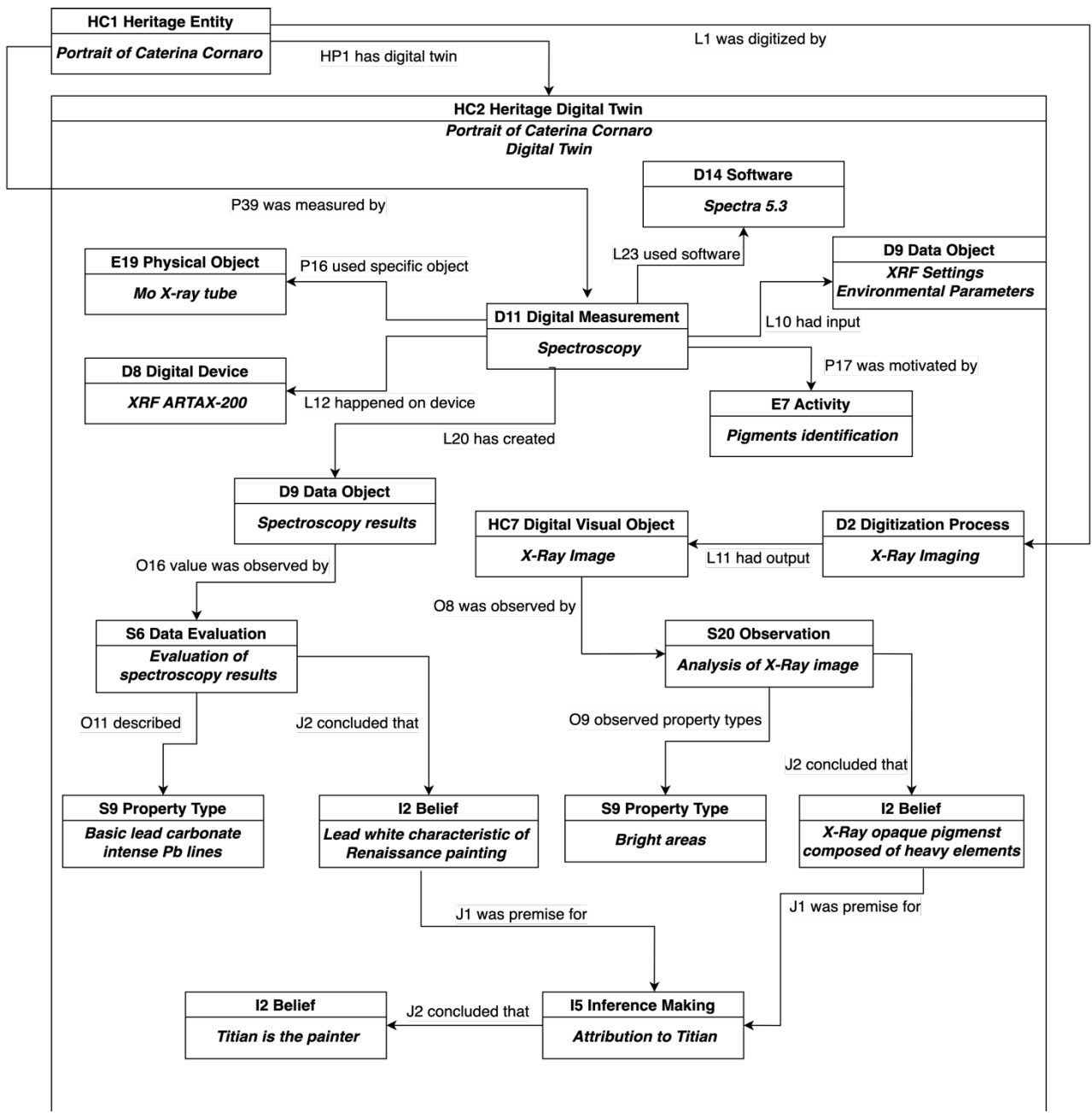

Figure 5. Modelling Heritage Science research activities and their results

## 8.2 The monastery of St. John Lampadistis in Kalopanayotis, Cyprus.

### 8.2.1 Tangible aspects

The monastery of St. John Lampadistis consists today of a group of buildings built and renovated in different periods. While its founding date is unclear, the *katholicon* (the monastery church), dedicated to Saint Herakleidios, born in the village of Lampadistis and the first bishop of Cyprus, is dated to the 11th century. Among the wall-paintings of the narthex an inscription, dated to the 15th century, describes this church as a "*katholiki*". While written sources indicate the monastery was in use until the 19th century, afterwards it has been used as the main church of the village. Apart from the complex of the three churches under one roof, a phenomenon unique for Cyprus, there are other monastic buildings including cells, auxiliary rooms and an oil press. One of the rooms is used today to house icons from the monastery as well as other churches of the village of Kalopanagiotis [31].

The main monastery church is a domed cross-in-square structure, dated to the 11th century. In the 12th century the vaulted chapel of St. John Lampadistis was added to the north of the first church, above the tomb of the Saint. Relics of the saint, kept and displayed in a rich reliquary, and his tomb are now incorporated in the church as part of this chapel. This second chapel collapsed and was almost entirely rebuilt in the 18th century. In the middle of the 15th century a common narthex was built to the west of the two churches.

During the second half of the 15th century a vaulted chapel was added to the north of that of St. John. It became known as the 'Latin chapel' because of the assumption that it was built for the Latins (Catholics) and indicating the possible co-existence of the two rites in the same church, in line with the atmosphere of tolerance which prevailed in Cyprus after the Council of Florence (1439). Sometime between the 15th and the beginnings of the 18th century), a timber roof covered with flat hooked tiles sheltered the entire roof complex. As a result of its tripartite character, the building acquired an external image of a large building covered with a timber roof.

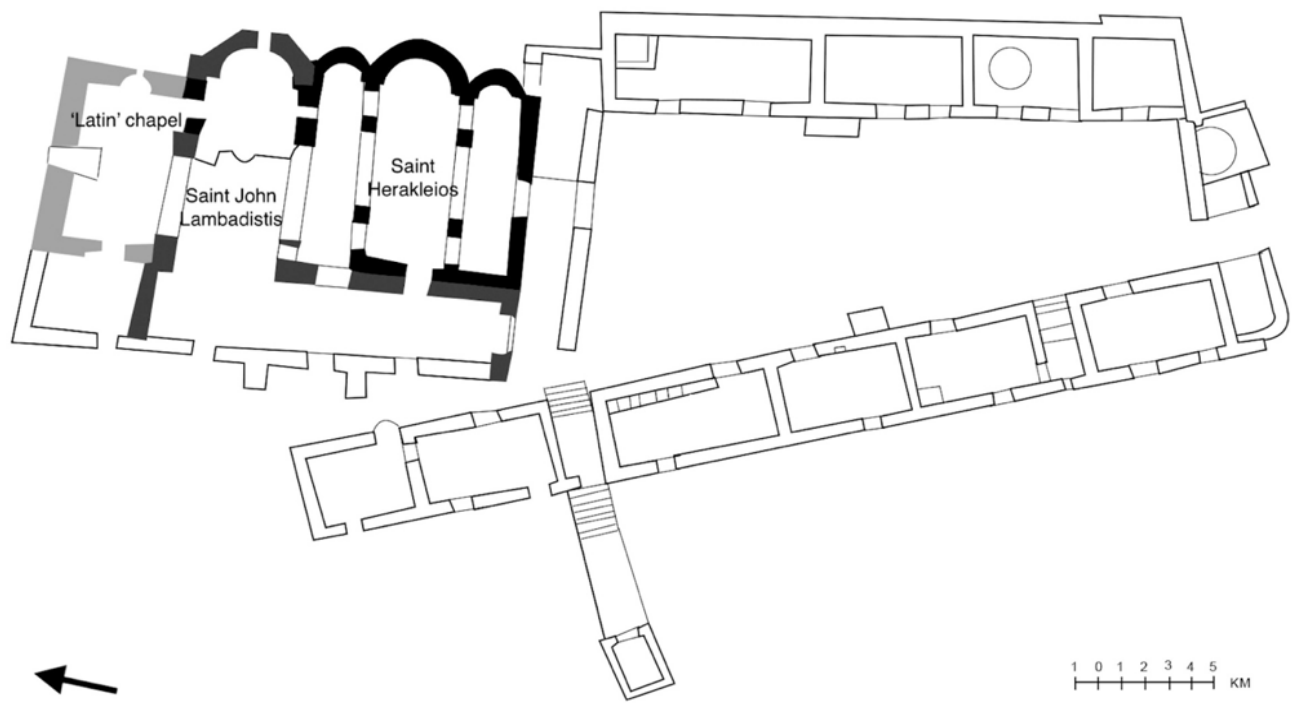

Figure 6. The plan of St. John Lampadistis complex (after [32], Figure 4)

The wall-paintings of the church are primarily from the 11th to the 15th centuries. The apse of the church, as well as some other parts, preserve fragmentary scenes dated to the 11th and 12th century. The rest of the church was painted in the 13th and 14th century. The narthex was decorated in a later period and is the work of an artist from Constantinople, who fled to Cyprus after the fall in 1453. These wall-paintings follow the trends of the Byzantine capital. The 15th century frescoes of the 'Latin' chapel display Byzantine and Italian Renaissance elements. A wooden templon screen, the oldest in Cyprus and dated to the 13th-14th century, is painted with decoration imitating coats-of-arms.

### 8.2.2 Intangible aspects
*The life of St. John Lampadistis*
A now lost 1640 manuscript written by a priest named Savvas from the village of Agios Theodoros Agrou and later copied by monk named Kirililos, of the Stavrovouni Monastery in 1903 narrates the life of St. John Lampadistis and the prayer said in church on his feast day on 4th of October. St. John was born in the village of Lampadistis, a now extinct village presumably located somewhere between the modern villages of Galata and Kakopetria, being the last offspring of the priest's village, Papa Kyriakos and his wife, Anna. His birthdate is unknown, probably sometime in the early 17th century. Early in his life, he was sent to learn to read and write through the study of the Holy Scriptures, the child showing great aptitudes. At the time of his adulthood, a marriage was arranged for him, but the parents of his future wife served him poisoned fish, causing him to lose his eyesight. Now being unsuitable to marriage, St. John turned to spiritual life and spent his days in prayer. After 12 years of living as a blind person, he died. Soon people reported light beaming from his grave. Word spread and people known to be possessed came seeking John's body to pray to it and be cured. Following pressure from the people, his father agreed to open his son's tomb and found his relic's, while his heart was preserved "like a dry fig". Hence, John's remains were deposited in the chapel of St. Herakleidios at Kalopanayiotis. The many miracles attributed to John earned him the status of a saint. His father, hearing a voice telling him to build a church dedicated to his son, which he did and the church of St. john Lampadistis was built adjacent to the one of St. Herakleidios. St. john's reputation as a miracle maker expanded and the inhabitants of the island decided to view him as their guardian saint.

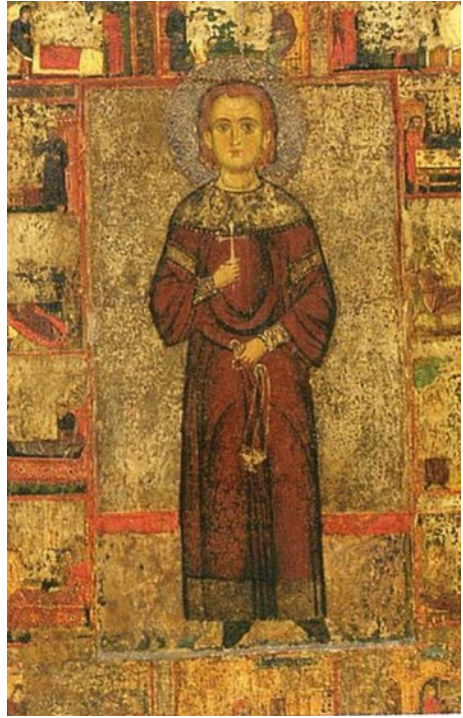

Figure 7. Vita icon depicting St. John and central episodes from his life

*Pilgrims and their testimonies - the Karamanlides*
The Karamanlika are a Turkish-speaking population of Greek Ortodox rite originally from Asia minor. During the 18th century a group of them, while on pilgrimage to the Holy Land, passed by the monastery of St. John Lampadistis and left several written testimonies of their passage on the church's wall close to its reliquary, the earliest being dated to 1749 and the latest to 1880, with a highest frequency between 1770 and 1780.

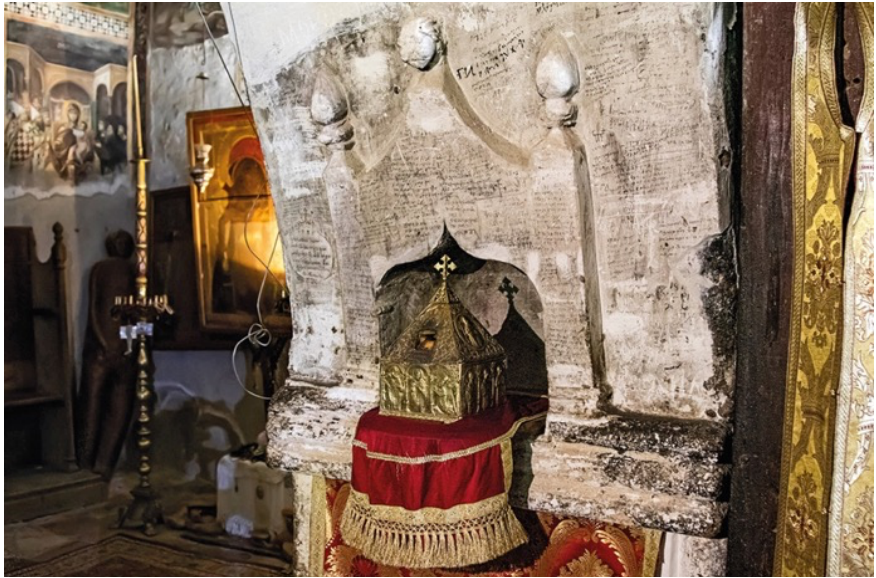

Figure 8. The Saint John Lampadistis reliquary, showing graffiti carved on the wall by Karamanlika pilgrims

### 8.2.3 Digital models
The 3D model of the monastery and related historical imaging documentation are available here:
https://modelier.us.aldryn.io/models/d10a14ee-4192-45b6-b54b-8d11ecd6f70e/v2/embed/
and a 360 panoramic walk-through is available here:
https://dioptra.cyi.ac.cy/sites/360/lampadistis/
part of a broader online-accessible database of the Troodos complex monasteries registered as World Heritage Sites http://ihat.cyi.ac.cy/?q=Collection

### 8.2.4 Ontological modelling examples
From the ontological point of view, the information about the structure of the monastery of St. John Lampadistis can be modelled by combining the classes and properties of the HDT Ontology with those of CIDOC CRM and CRMba. The monastery (HC1) is in fact made up of various buildings (P46 → B1) including the central church (*katholicon*) where the frescoes are found. Each building can in turn contain several building sections (BP1 → B2) clearly identifiable by their morphology, and of each of them it is possible to specify and describe the construction phases (P108 -> E12) and the relative dating (P4 → E52). As far as the frescoes on the walls are concerned, it is possible to identify those pertaining to each section of the building (P56 → E26) so that even for them the phases of realization (P108 → E12) and the dating (P4 → E52) can be clearly distinguished. Figure 9 illustrates the case of the modelling of the Latin Chapel and its frescoes, highlighting, through the distinction between the two different production activities (E12 → P4 → E52), how the painting of this section of the building is coeval with the construction of the chapel itself (both belonging to the 15th century, in this case). The monastery is also visually represented through various digital technologies (3D models, 360 panoramic views, etc.) which can be perfectly represented through our modelling tools either with (P129 → HC8) as in Figure 9, or with (L1/L11 → HC8) when describing the digitisation process is needed.

The HDT Ontology and the CIDOC CRM ecosystem can be also employed to model the intangible aspects of the monastery, represented by its history (HP5 → HC4 and HP2 → Narrative) and linked in particular to the life of St. John Lampadistis, narrated by the priest Savvas (HP4 → Narration and HP8 → E31) in a manuscript (P128 → E18) that is now lost (P44 → E34). It is also possible to model the information related to the graffiti engraved on the walls of the monastery church by the Karamanlika pilgrims (P56 → E26), to specify their approximate dating (the 18th century in this case)(P4 → E52), and to link the images that depict them (P138 → E36).

Finally, the complex of information described above and their mutual relationships form the elements that make up the Heritage Digital Twin of the monastery of St. John Lampadistis. The HDT Ontology establishes this fundamental relationship via the property HP1 has digital twin, used to bind each Heritage Entity (HC1) to its digital representation (HC2 Heritage Digital Twin).

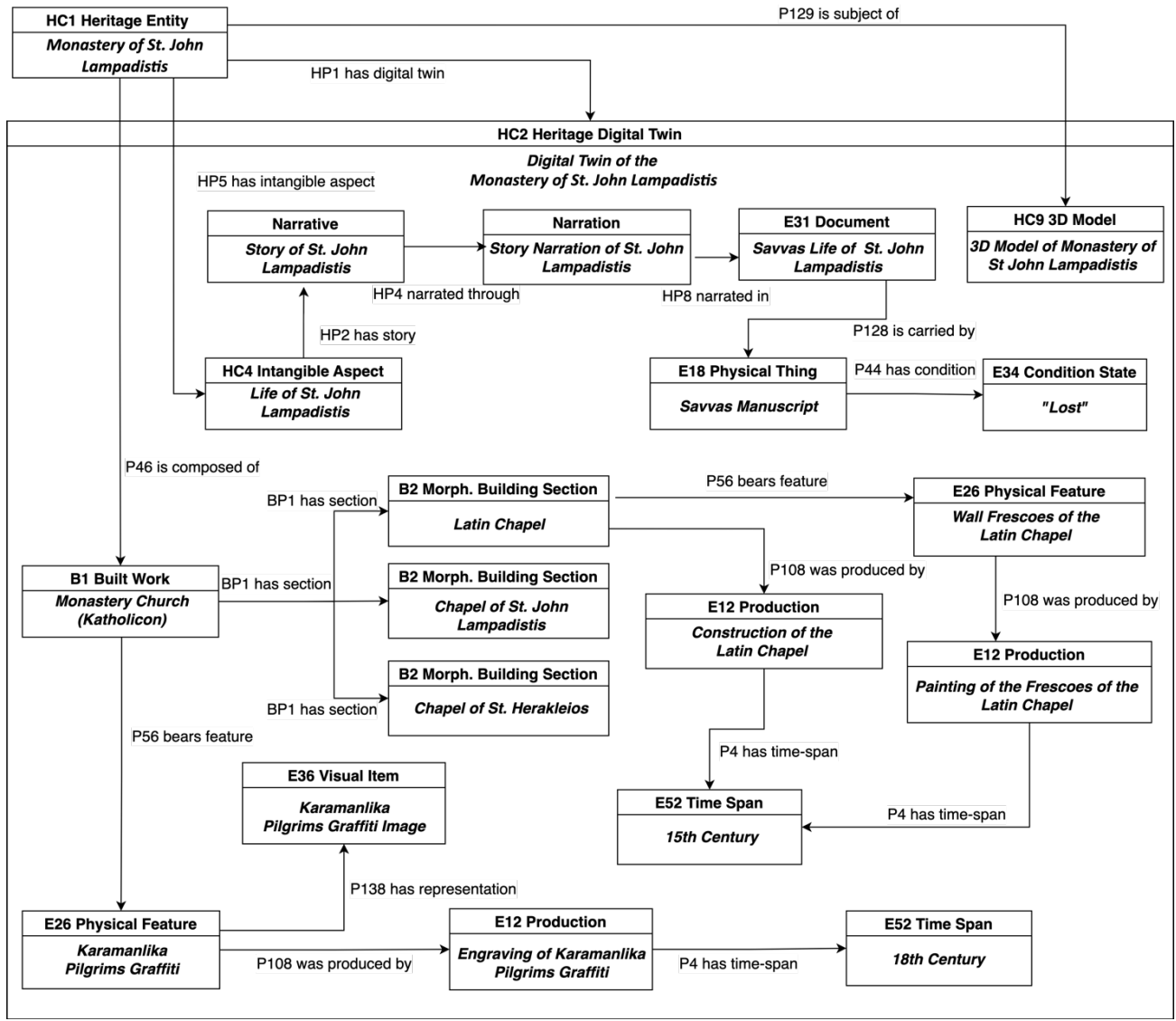

Figure 9: The ontological model of (part of) the Heritage Digital Twin of St. John Lampadistis Monastery

## 8.3 Final considerations on the examples

The patterns built by combining the different extensions of the CIDOC CRM and our new classes and properties are used to model different situations and scenarios. They can be replicated to describe all the other observations, analyses, hypotheses, and any inferences and conclusions that can be drawn from them. The conclusions in turn can serve to support new hypotheses and theories, the graph can be extended without any limitation.

Finally, as seen in the diagrams, all the semantic information acquired about the Heritage Entity and modelled through the ontology eventually becomes part of the Digital Twin (HC2), in which all the knowledge about the object is organized.

## 9. Conclusions and further work

This paper is the outcome of a research activity which is still ongoing. We decided to publish it as it is now, in our opinion a stable and coherent version, to open a discussion and improve it with the contributions of the users' community. We believe that the proposed solution reconciles the different and sometimes incompatible approaches summarily sketched in section 3 and enables an approach a la carte according to different user needs while preserving overall interoperability. Existing documentation may be easily mapped to the HDT ontology and different systems may be integrated into it, combining the existing information under the HDT overarching umbrella. Doing so will enhance the interoperability of different systems and enable the creation of a well-organized data space to support the work by researchers, heritage professionals and the public. Moreover, the HDT is ready for simulation – the second step of digital twin systems – such as answering to "what if" questions as well as causing automatic reactions to actual external events. The CRM community (CRM SIG) is currently working on a new extension for representing not only past activities but also future ones, CRMact[14], that could be the way towards modelling such dynamics.

Our plans for future work include polishing the HDT ontology with the suggestions coming from the community. In parallel, we will start mapping existing schemas, such as the Europeana EDM ontology[15] or those based on HBIM or 3D annotations, on the HDT ontology adding specific classes and properties if needed, to ultimately create a framework where everybody can recognize their own documentation scheme but can also interoperate with those created by others.

We believe that this objective is a necessary condition to support collaborative scenarios on digital cultural heritage as those envisaged by the EU strategies on digitization.

**Grant information**
The research leading to the present paper was partially funded by the following grants:
ARIADNEplus: European Commission programme H2020 (INFRAIA-01-2018-2019) Grant no. 823914
4CH: European Commission programme H2020 (DT-TRANSFORMATIONS-20-2020) Grant no. 101004468

---

[14] https://cidoc-crm.org/crmact/node/8732
[15] https://pro.europeana.eu/page/edm-documentation